# STEP BUNCHING IN CONSERVED SYSTEMS: SCALING AND UNIVERSALITY

B. Ranguelov[1], V. Tonchev[1,*], C. Misbah[2]
[1]Institute of Physical Chemistry, Bulgarian Academy of Sciences, 1113 Sofia, Bulgaria
[2]Laboratoire de Spectrométrie Physique, CNRS/Université Joseph Fourier, Saint-Martin d'Hères

**Abstract**. We study the step bunching process in three different 1D step flow models and obtain scaling relations for the step bunches formed in the long times limit. The first one was introduced by S.Stoyanov [Jap. J.Appl. Phys. **29**, (1990) L659] as the simplest 'realistic' model of step bunching due to drift of the adatoms. Here we show that it could lead to (at least) two different types of step bunching, depending on the magnitude of the drift. The other two models are *minimal models*: the equations for step velocity are constructed *ad hoc* from two terms with opposite effects - destabilizing and, respectively, stabilizing the regular step train.

**Introduction**. The evaporation and growth of the vicinal crystal surfaces are mediated by the existing monoatomic steps usually uniformly distributed over the surface. Since the seminal work of Burton, Cabrera and Frank (BCF) [1] the step flow growth mode is intensively studied both theoretically and experimentally and one of the main reasons is that the Molecular Beam Epitaxy (MBE) and related techniques are the basis of the modern semiconductor technology. Since the discovery of step bunching during high temperature DC heating of Si(111) vicinals by Latyshev et al. [2] there is an increasing interest in modeling the unstable step motion on the crystal surface. Initially the studies were focused on predicting the onset of the instability [3,4]. Later the step bunching process was studied in the long times limit and a scaling relation for the minimal interstep distance in the bunch was obtained [5] both analytically and numerically and confirmed in sublimation experiments[6,7]. Following the hypothesis of universality classes in bunching by Pimpinelli, Tonchev, Videcoq and Vladimirova (PTVV) [8] a rather complete theory of the step bunching in 1D due to 'classical' mechanisms of destabilization - Ehrlich-Schwoebel (ES) effect and electromigration - is proposed [9] but some more questions remain open. Among them are: (i) to which universality class belong the models and experimental systems studied before the work of PTVV [8]; (ii) could one obtain different scaling regimes/universality classes with only changing the parameters in a discrete step flow model similarly to the analytical results in [9]; (iii) could one build a discrete analog to the based on a continuum equation classification scheme of PTVV [8]; (iv) how many universality classes in bunching exist. Before looking for some of the answers it should be mentioned that step bunches formed on Si(001) and Si(111) vicinals were used as pre-patterned substrates with possible applications in the nano-device production technology [10,11,12]. Thus step bunching models are not only a toy for the theorists but could serve to provide a useful guidance for the practitioners.

**The Models.** (i) The model of **step bunching due to drift of the adatoms** (DA) is first formulated by S. Stoyanov [3]. Since it was studied before the PTVV classification scheme was proposed [8] and studied numerically [13] the scaling exponents obtained are not enough to attribute the model to a specific universality class and this is the initial reason to revisit the model. In what follows we briefly remind the main features of the model and the interested reader is suggested to revise the paper of Sato and Uwaha [13] for more details. The adatom migration is restricted to a single terrace from the vicinal surface bounded by the two adjacent steps. The stationary diffusion equation of the model contains only two terms:

$$D_s \frac{\partial^2 C}{\partial x^2} - v \frac{\partial C}{\partial x} = 0 \qquad (1)$$

where $D_s$ is the coefficient of surface diffusion, $C(x)$ is the concentration of mobile atoms and $v$ is the drift velocity. The integration constants are found using two boundary conditions on the two adjacent (impermeable) steps denoted with (-) and, respectively, (+) sign for the left and the right step:

$$\pm d_l \left( \frac{\partial C}{\partial x} - FC \right) = C - C_\pm^e \qquad (2)$$

where $F=v/D_s$, $d_l=D_s/K$ is the kinetic length[14], $K$ being the step kinetic coefficient, and the effect of the step-step repulsions (described by the power law $U=A/l^n$) is incorporated[5] in the equilibrium concentrations $C_\pm^e$. To be more specific, the interactions dependent equilibrium concentration $C_i^e$ in the vicinity of the $i$-th step is:

$$C_i^e = C_0^e \left[ 1 + \frac{n \Omega A}{kT} \left( \frac{1}{\Delta x_{i+1}^{n+1}} - \frac{1}{\Delta x_i^{n+1}} \right) \right] \qquad (3)$$

where $\Omega$ is the crystal surface area per atom, $C_0^e$ is the equilibrium concentration around a step from the vicinal step configuration and $\Delta x_{i+1} = (x_{i+1} - x_i)$ is the width of the terrace between $i$-th and $i+1$-st step. In fact, the step velocity $V_i$ of the $i$-th step placed in $x_i$ is obtained from the contributions of the atom concentrations on the both terraces adjacent to the step:

$$V_i \equiv \frac{dx_i}{dt} = K\Omega \{[C_-(x_i) - C_-^e] + [C_+(x_i) - C_+^e]\} \qquad (4)$$

One can easily estimate from Eqs.3 and 4 the influence of the step-step repulsions on the step velocity assuming that the $i$-th step from a moving in $+x$ direction step train is disturbed towards the $i+1$-st. This causes increase of $\Delta x_i$ (the width of the terrace behind the step) and decrease of $\Delta x_{i+1}$. As a result, the equilibrium concentration $C_i^e$ will increase leading to a decrease of step velocity $V_i$. The opposite effect will be caused if the $i$-th will slow down due to some disturbance – the interstep repulsion will cause an increase of the step velocity.

After solving the Eq. 1 with the boundary conditions, Eqs.2, one obtains (linearized) equatons for the dimensionless step velocity $v_i$:

$$\frac{1}{\Omega C_0^e} v_i = \frac{1}{\Omega C_0^e} \frac{d\xi_i}{d\tau} =$$

$$\frac{(l_0/l)^{n+1}\left[\frac{1}{\Delta\xi_{i+2}^{n+1}} - \frac{2}{\Delta\xi_{i+1}^{n+1}} + \frac{1}{\Delta\xi_i^{n+1}}\right] - f\Delta\xi_{i+1}\left[1 + (l_0/l)^{n+1}\left(\frac{1}{\Delta\xi_{i+1}^{n+1}} - \frac{1}{\Delta\xi_i^{n+1}}\right)\right]}{2(d_l/l) + \Delta\xi_{i+1}[f(d_l/l) + 1]} -$$

$$\frac{(l_0/l)^{n+1}\left[\frac{1}{\Delta\xi_{i+1}^{n+1}} - \frac{2}{\Delta\xi_i^{n+1}} + \frac{1}{\Delta\xi_{i-1}^{n+1}}\right] - f\Delta\xi_i\left[1 + (l_0/l)^{n+1}\left(\frac{1}{\Delta\xi_i^{n+1}} - \frac{1}{\Delta\xi_{i-1}^{n+1}}\right)\right]}{2(d_l/l) + \Delta\xi_i[f(d_l/l) + 1]} \quad (5)$$

where $\xi_i = x_i/l$, $\tau = tl^2/D_s$, $f = Fl$, $l_0 = (n\Omega A/kT)^{1/(n+1)}$, and $l$ is the initial vicinal distance. The main reason we give the explicit form of Eq.5 is to demonstrate that even the simplest 'realistic' model results in quite sophisticated equations for the step motion. This is the reason to look for simpler models containing the minimal prerequisites leading to step bunching [18] (see also [20] where minimal models are introduced from a somewhat different perspective) and to try to understand better their behavior before turning again to the 'realistic' ones. Note that the continuum equation used by PTVV [8] to advance the universality classes hypothesis is also a minimal one. The models are constructed *ad hoc*. The equations for the step velocity are defined using analogies with the 'realistic' models and consist of two parts: destabilizing and stabilizing. The former is that leads to bunching of the steps while the later prevents the steps from overlapping.

(ii) The first of the minimal models we introduce is called **Minimal Model I** (MMI). The equations for the step velocity in MMI are defined as:

$$\frac{dx_i}{dt} = K\left[(x_i - x_{i-1})^r + b(x_{i+1} - x_i)^r\right] - A\left[\frac{1}{(x_{i+1} - x_i)^n} - \frac{1}{(x_i - x_{i-1})^n}\right] \quad (6)$$

Both terms on the rhs of Eq.6 contain only the widths of the two adjacent to the step terraces. The comparison with Eq.5 shows that it contains the widths of four terraces – nearest and next nearest neighbors of the step. The first term on the rhs of Eq.6 may lead to destabilization of the regular step train while the second term always stabilizes the uniform distribution of the steps and this is a difference from the case of infinite ES effect in evaporation where the interstep repulsion vanishes from the equations of step velocity[9]. Our MMI resembles the continuum equation of PTVV with $r$ being an analog to $\rho$, the power of the local slope in the destabilizing term of this continuum equation. Further, we carry out a linear stability analysis assuming first that the steps are moving in a regular step train with all distances equal to $l$. In this case their velocity is:

$$V_0 = Kl^r(1 + b) \quad (7)$$

Now assume that the position of the $i$-th step is disturbed by a small value $\delta l$ changing the widths of the neighboring terraces to: $x_i - x_{i-1} = l + \delta l$ and $x_{i+1} - x_i = l - \delta l$. The velocity of the step is then:

$$V_i = Kl^r(1 + b) + \delta l\left[Krl^{r-1}(1 - b) - \frac{2An}{l^{n+1}}\right] = V_0 + \delta V \quad (8)$$

where only the terms linear in $\delta l$ are kept. Thus the change of the velocity is:

$$\delta V = \delta l\left[Krl^{r-1}(1 - b) - \frac{2An}{l^{n+1}}\right]$$

The movement of the initially regular step train is unstable when positive $\delta l$ leads to a positive $\delta V$ and vice versa:

$$Krl^{r-1}(1-b) - \frac{2An}{l^{n+1}} > 0 \Rightarrow \frac{A}{K}\frac{1}{l^{n+r}(1-b)} < \frac{r}{2n} \qquad (9)$$

Thus, we obtained stability condition, Eq.9, for the parameters of MMI to be used in the studies of the bunching in this model. Also, it is seen that when $b=-1$ $V_0=0$, the position of mass centre remains unchanged while step movement still could be unstable leading eventually to bunching and in this case the bunching process is called 'hierarchical' or 'conserved'.

(iii) The idea for the next model, the **Minimal Model II** (MMII), comes from the previous one. If the second term in the rhs of Eq.6 stabilizes the uniform step distribution then the same term with opposite sign could be used as a destabilizing one:

$$\frac{dx_i}{dt} = K\left(\frac{1}{(x_{i+1}-x_i)^r} - \frac{1}{(x_i-x_{i-1})^r}\right) - A\left(\frac{1}{(x_{i+1}-x_i)^n} - \frac{1}{(x_i-x_{i-1})^n}\right) \qquad (10)$$

The difference between the two terms is only that the terrace widths are raised to different powers, $r$ and $n$ respectively, in order to study their effect independently. The instability condition for this model is:

$$\frac{A}{K}\frac{1}{l^{n-r}} < \frac{r}{n} \qquad (11)$$

In fact, MMII could be obtained from MMI with $b=-1$ and $r$ negative, it is also seen from the stability conditions, but we separated both models for better illustration of the results.

**Results and Discussion**. The equations of step motion, Eqs.5, 6 and 10, are integrated numerically using fourth order Runge-Kuta procedure[15]. Thus we follow qualitatively the step trajectories and surface profile evolution but also obtain scaling relations for the step bunches. Definitions of the quantities that are monitored during the calculations are shown on Fig.1. First we study the model of bunching due to DA. Step trajectories, surface profile and surface slope are shown on Fig.2 and Fig.3 for two different values of the drift. There is a qualitative difference seen on the plot of the step trajectories. While for the smaller value of the drift there are no steps crossing the terrace (Fig.2a) for the larger value of the drift the number of steps crossing the terraces increases with the development of the instability (Fig3a). It is not clear how this qualitative observation is connected with the difference in the scaling exponents reported below. Another important observation is that it is impossible to distinguish both regimes looking at the shape of the bunches and especially at the slope of the surface as shown on Figs 2b and 3b. There is expectation that a different matching of the bunches with the terraces, smooth or abrupt [16], would lead to a different scaling but we are unable to determine to which of the two limiting cases belong the bunches studied. What is clear is that the difference in the magnitude of the drift leads to two different scaling regimes as shown on Fig.4 and Fig.5. The size scaling exponents, as shown on Fig.4, obtained for the two different values of the drift, represent two archetypical cases (note that for the studies of this model we use only the canonical power of the interstep repulsion $n=2$). Thus for the size-scaling of the minimal interstep distance, $l_{min} \sim N^\gamma$, we obtain $\gamma=2/3$ and $1/2$, respectively and for the scaling of the bunch width with the bunch height, $W \sim H^{1/\alpha}$, we obtain $\alpha=3$ and $2$, respectively. The value $\gamma=2/3$ was obtained both theoretically, numerically [5,9] and experimentally[6,7], as well as the value $\alpha=3$ obtained numerically[17] and theoretically [9]. Recently, the values $\gamma=1/2$, $\alpha=2$ were also obtained theoretically [9] but no numerical or experimental confirmation was provided so far. Unfortunately, it is impossible from the collected time-scaling data, Fig.5, to attribute observed regimes of bunching to any of the known universality classes [8,9] and hence more systematic studies are needed to reveal the complete universality picture of the model. Further, we briefly summarize our results on minimal models. On Fig. 6a are shown step trajectories obtained in the MMI, while on Fig.6b are shown surface profile and the surface slope obtained in MMII. Note that MMII shows a unique - the interstep distances in the bunch remain constant despite the increase of the number of steps in a bunch. This was not observed in any other step bunching model. The scaling relations we obtain for the both models are:

$$L_b \sim aH^{\frac{n-r}{n+r-1}} \qquad H \sim t \qquad L_b \sim t^{\frac{n-1}{n}} \qquad \text{(MMI)}$$

$$Lb \sim a^{\frac{1}{n-r}}H \qquad H \sim t^{\frac{1}{r+2}} \qquad \text{(MMII)}$$

Both models generate new universality classes in bunching summarized in the table of universality classes in [19].

**Acknowledgements.** This study is financially supported by Grant F-1413/2004 from the Bulgarian National Science Fund. VT acknowledges hospitality and stimulating working conditions at Physics Department of the Joseph Fourier University, Grenoble, France.

**References**

1. Burton, W. K. et al., Phil.Trans Royal Soc. London **243**, 866 (1951) 299.
2. Latyshev, A. V. et al., Surf. Sci. **213**, 1 (1989) 157.
3. Stoyanov, S., Jap. J. Appl. Phys. **29**, 4 (1990) L659.
4. Stoyanov, S., Jap. J. Appl. Phys **30**, 1 (1991) 1.
5. Stoyanov, S. and Tonchev, V., Phys. Rev. B **58**, 3 (1998) 1590.
6. Fujita, K. et al., Phys. Rev. B **60**, 23 (1999) 16006.
7. Homma, Y. and Aizawa, N., Phys. Rev. B **62**, 12 (2000) 8323.
8. Pimpinelli, A. et al, Phys. Rev. Lett. **88**, 20 (2002) 206103.
9. Krug, J. et al. Phys. Rev. B **71**, 045412 (2005).
10. Sgarlata, A., et al., Appl. Surf. Sci. **83**, 19 (2003) 4002.
11. Patella, F., et al. J. Phys.: Cond. Matt.**16**, (2004) S1503.
12. Lichtenberger, H. et al., Appl. Phys. Lett. **86**, 131919 (2005).
13. Sato, M. and Uwaha, M., Surf. Sci. **442**, (1999) 318.
14. Pimpinelli, A., et al., J. Phys.: Cond. Matt. **6**, 14 (1994) 2661.
15. Press, W. H., Teukolski, S. A., Vetterling, W. T., and Flannery, B. P., Numerical Recipes in Fortran 77: The Art of Scientific Computing (Cambridge University Press, Cambirdge, 1997)
16. Pierre-Louis, O. et al., J. Cryst. Growth **275**, (2005) 56.
17. Stoyanov, S. et al., Surf. Sci. **465**, (2000) 227.
18. Tonchev, V., 2002, unpublished.
19. Ranguelov, B. et al., CR de l'Acad. Bulg. Sci. **60,** 4 (2007) 389.
20. H. Sakaguchi and N.Fujimoto, Phys. Rev. E **68**, 056103 (2003).

**Figure Captions**

Figure 1: Definitions of the quantities characterizing a step bunched vicinal surface: $w$ is the terrace width, $\phi$ is the vicinal angle, $h_0$ is the height of the (monoatomic) step, $l$ is the initial vicinal distance, $L_b$ is the width of the bunch. Note the difference in defining bunch size $N$ (number of step in the bunch) and bunch height $H$ (the height excess over the vicinal one). $H=w\text{tg}\phi$ holds only when there are no steps on the terrace. As seen, despite the formation of the bunch, the surface preserves its macroscopic crystallographic orientation.

Figure 2. Step trajectories (a) and surface profile/surface slope (b) obtained for the model of step bunching due to **drift of the adatoms** for 'small' value of the drift $f=0.085$, other parameters shown on the plot.

Figure 3. Step trajectories (a) and surface profile/surface slope (b) obtained for the model of step bunching due to **drift of the adatoms** for a 'large' value of the drift $f=0.4$, , other parameters shown ob the plot.

Figure 4. Size-scaling for the model of step bunching due to **drift of the adatoms** in the case of 'small', $f=0.085$, (a) and 'large', $f=0.4$, (b) value of the drift, other parameters are the same.

Figure 5. Time-scaling for the model of step bunching due to **drift of the adatoms** in the case of 'small', $f=0.085$, (a) and 'large', $f=0.4$, (b) value of the drift, other parameters are the same.

Figure 6: Step trajectories obtained for MMI (a) and surface profile/surface slope for MMII (b).

Figure 1

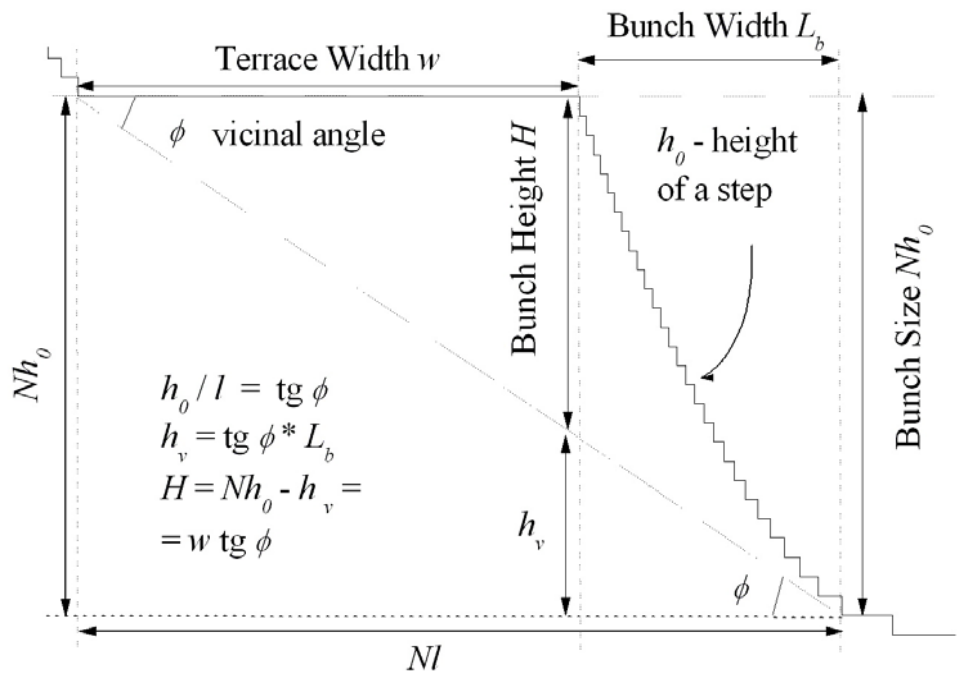

Figure 2a

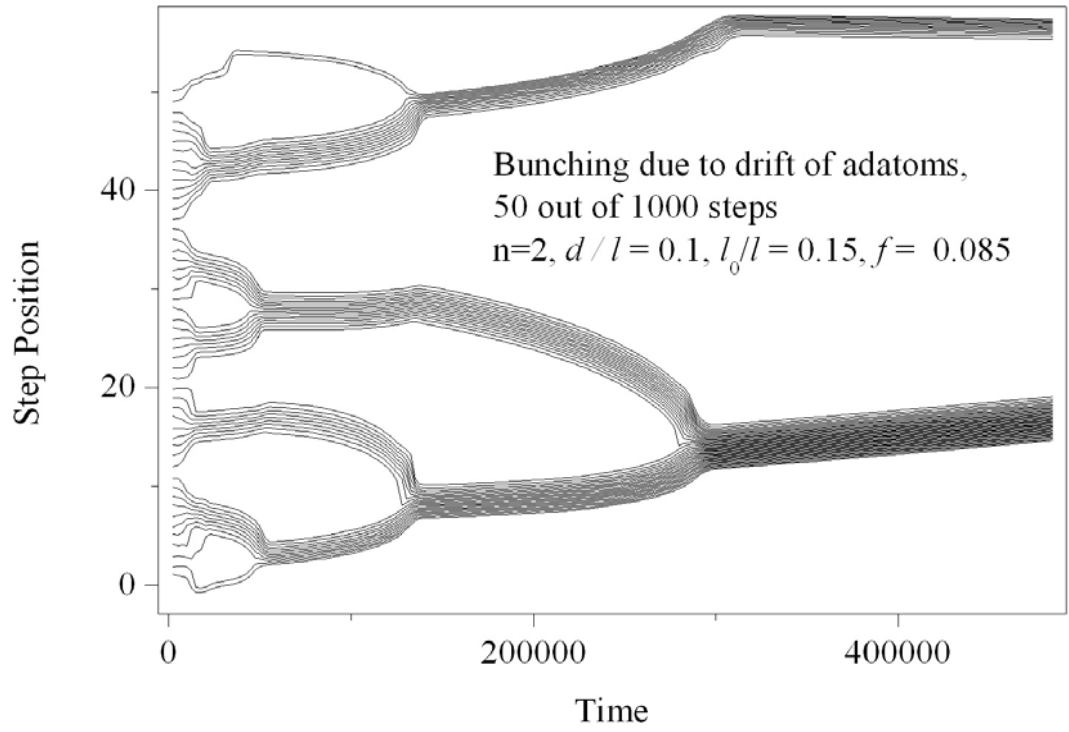

Figure 2b

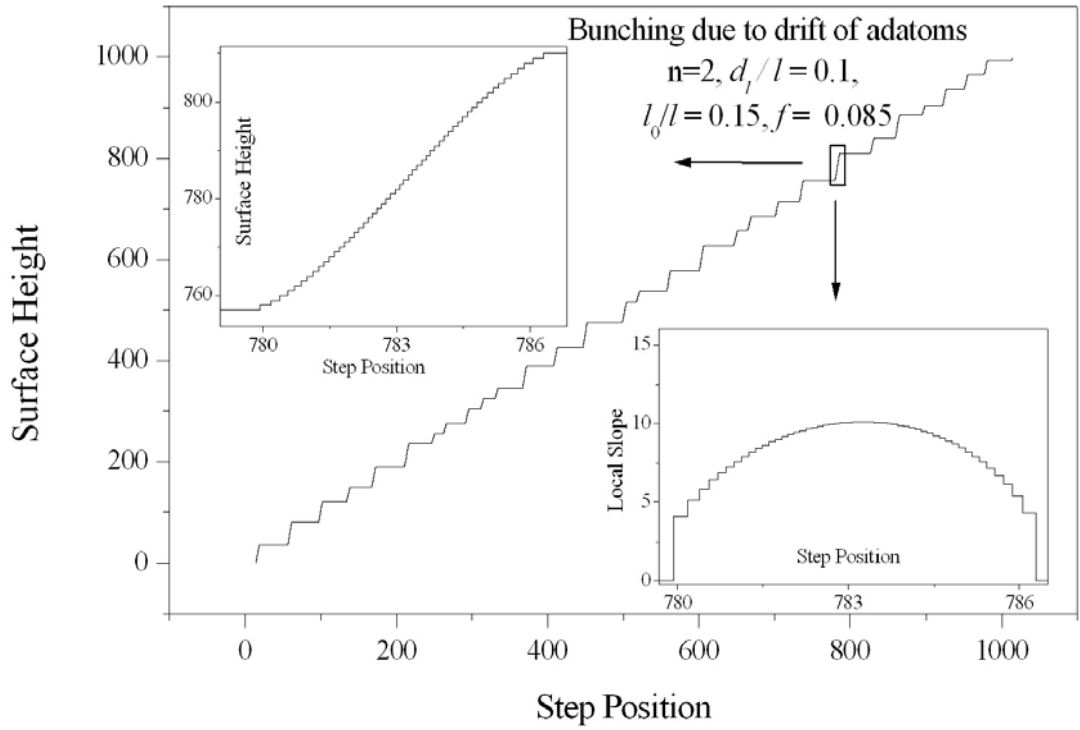

Figure 3a

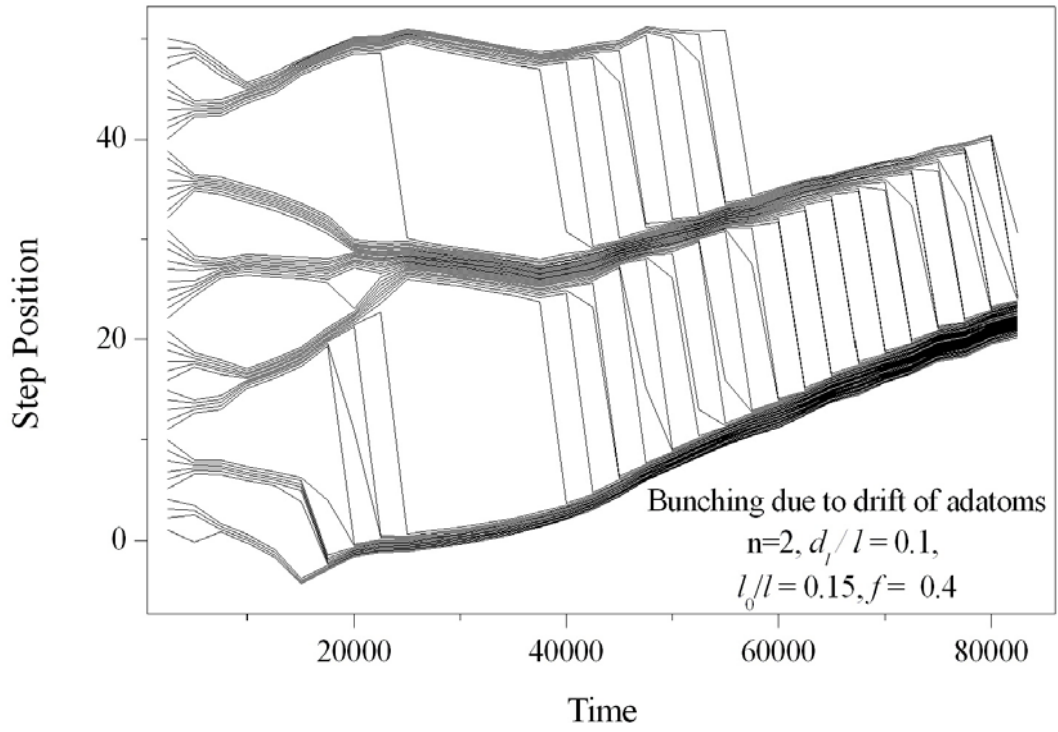

Figure 3b

[Figure: Plot of Surface Height vs Step Position showing bunching due to drift of adatoms with parameters $n=2$, $d_l/l = 0.1$, $l_0/l = 0.15$, $f = 0.4$. Inset plots show Surface Height and Local Slope vs Step Position near positions 408–410.]

Figure 4a

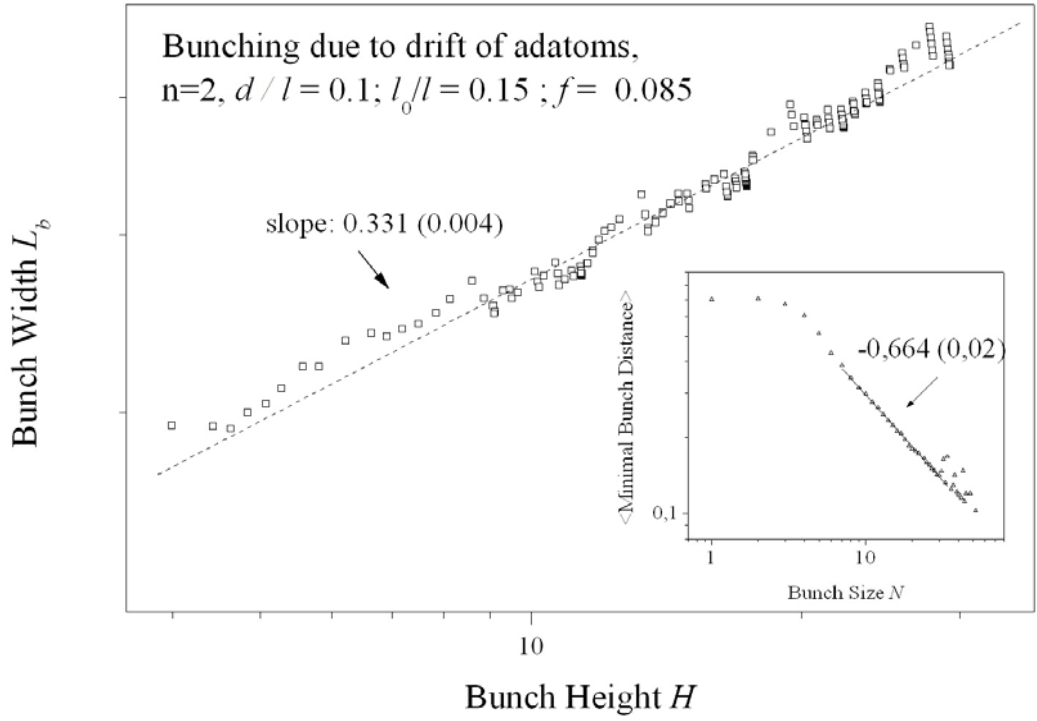

Figure 4b

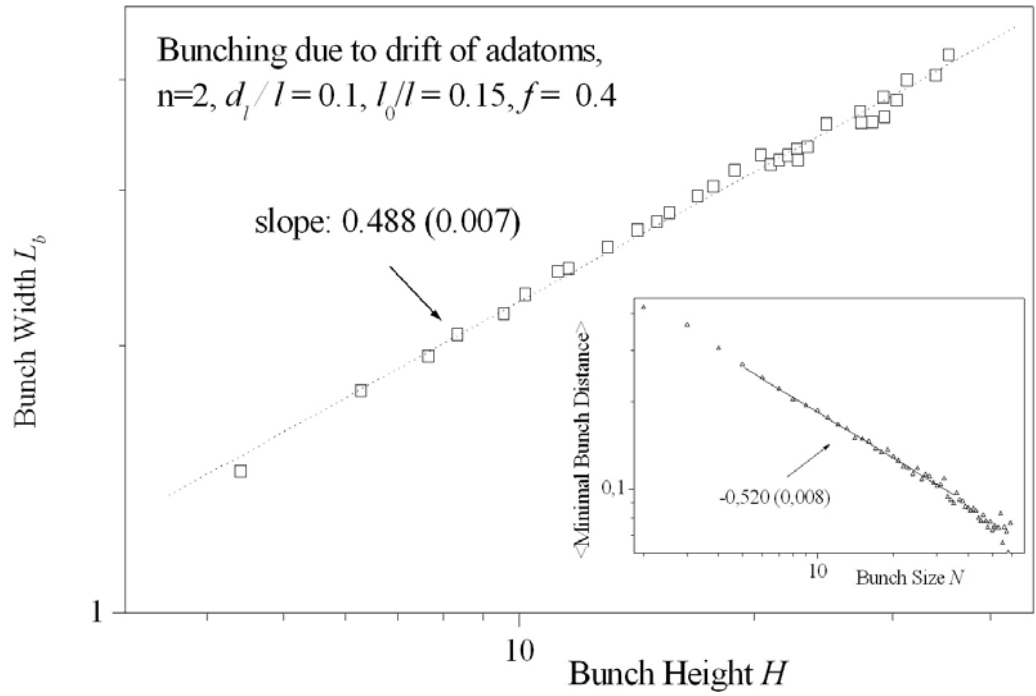

Figure 5a

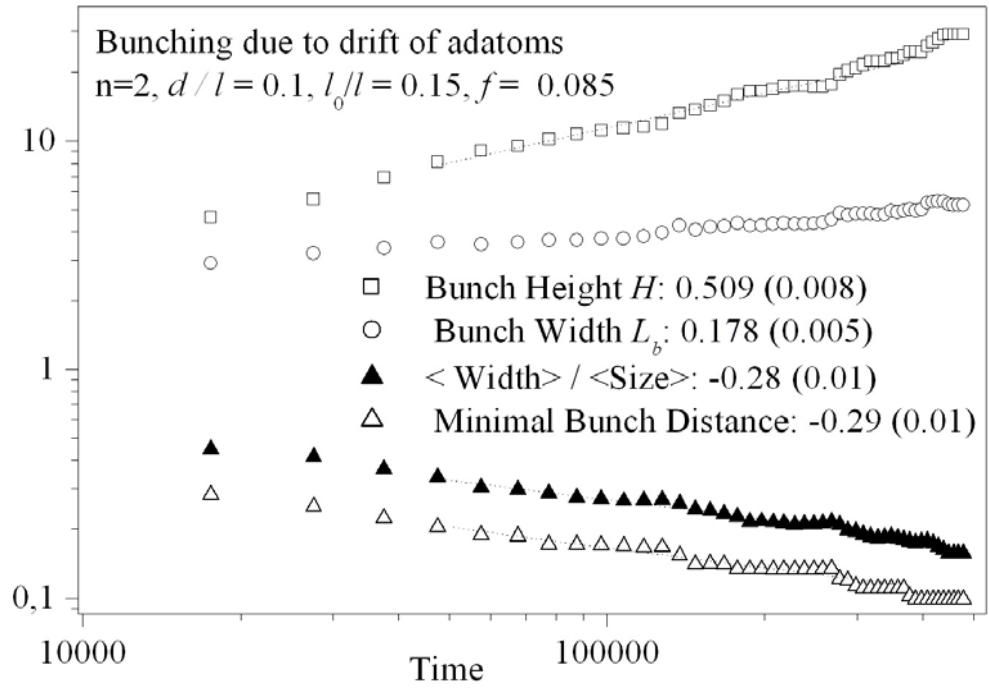

Figure 5b

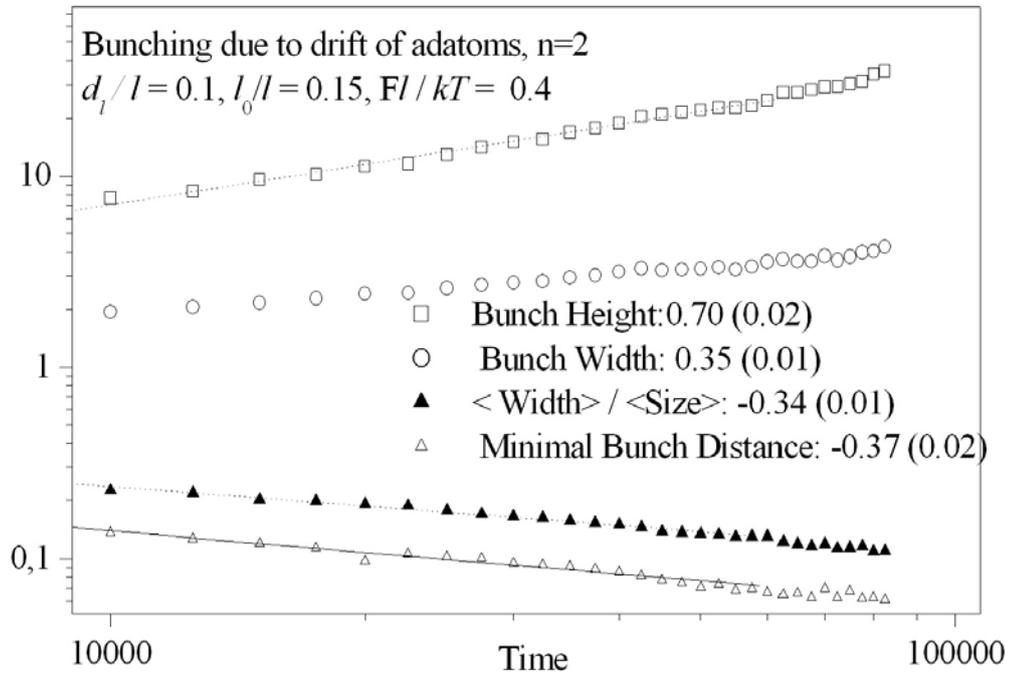

Figure 6a

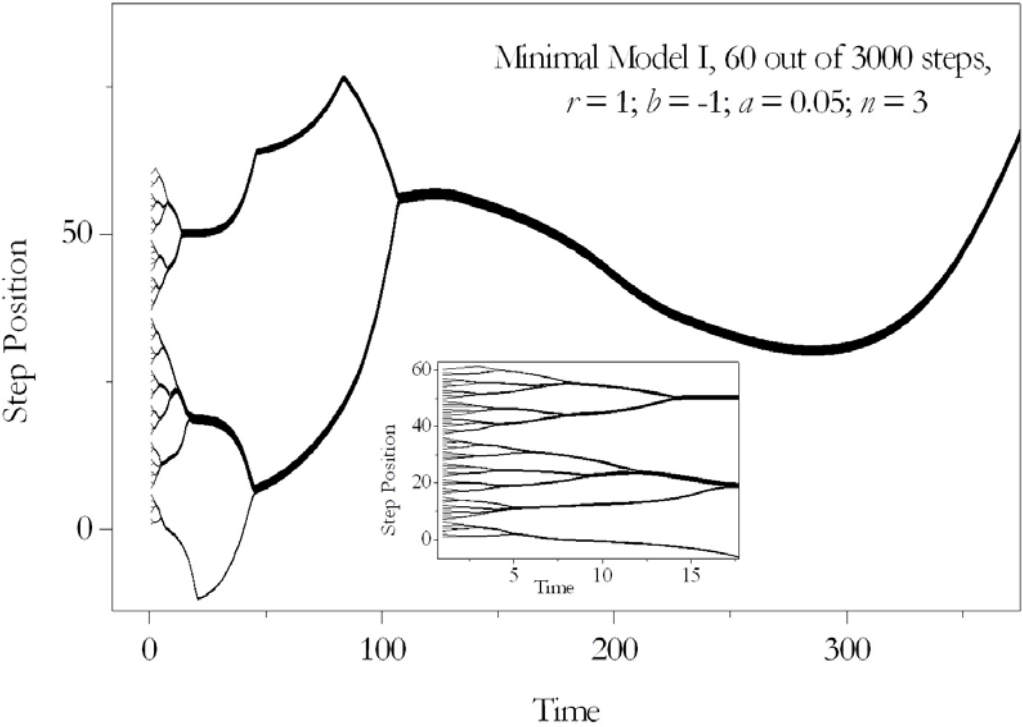

Figure 6b

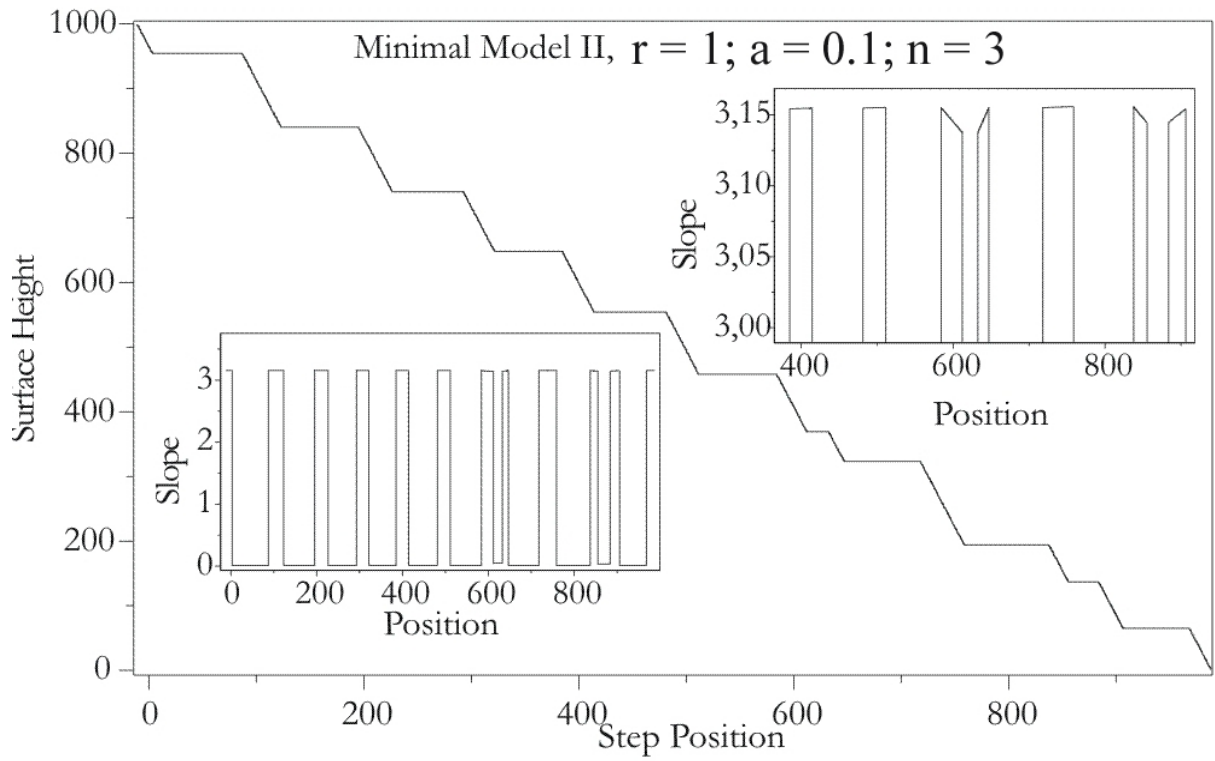